\documentclass[aps,prd,reprint,twocolumn,groupedaddress,superscriptaddress]{revtex4-2}
\usepackage{lipsum}
\usepackage{graphicx} 
\usepackage{dcolumn}  
\usepackage{bm}       
\usepackage{amssymb}   
\usepackage{amsmath}
\usepackage{xcolor}
\usepackage{amsmath}
\usepackage{amsfonts}
\usepackage{hyperref}
\hypersetup{colorlinks=true, linkcolor=blue, citecolor=blue, urlcolor=blue}
\usepackage{bm}
\usepackage{xcolor}

\begin{document}
\preprint{APS/123-QED}

\title{Extracting Transport Properties from the Heavy-Quark Potential with Neural Networks in a Holographic Model}

\author{Wen-Chao Dai}
\email{chenxun@usc.edu.cn}
\affiliation{School of Nuclear Science and Technology, University of South China, No. 28, West Changsheng Road, Hengyang City, Hunan Province, China.}

\author{Ou-Yang Luo}
\affiliation{School of Nuclear Science and Technology,
	University of South China, No. 28, West Changsheng Road,
	Hengyang City, Hunan Province, China.}
\author{Bing Chen}
\affiliation{School of Nuclear Science and Technology,
	University of South China, No. 28, West Changsheng Road,
	Hengyang City, Hunan Province, China.}

\author{Xun Chen}
\email{chenxunhep@qq.com}
\affiliation{School of Nuclear Science and Technology,
	University of South China, No. 28, West Changsheng Road,
	Hengyang City, Hunan Province, China.}
\affiliation{Key Laboratory of Advanced Nuclear Energy Design and Safety, Ministry of Education, Hengyang, 421001, China.}
\affiliation{INFN– Istituto Nazionale di Fisica Nucleare
	Sezione di Bari, Via Orabona 4, 70125 Bari, Italy.}
\date{\today}

\author{Xiao-Yan Zhu}
\email{xyzhu0128@163.com}
\affiliation{School of Mathematics and Physics,
	University of South China, No. 28, West Changsheng Road,
	Hengyang City, Hunan Province, China.}

\date{\today}

\author{Xiao-Hua Li}
\email{lixiaohuaphysics@126.com}
\affiliation{School of Nuclear Science and Technology,
	University of South China, No. 28, West Changsheng Road,
	Hengyang City, Hunan Province, China.}
\affiliation{Key Laboratory of Advanced Nuclear Energy Design and Safety, Ministry of Education, Hengyang, 421001, China.}
\date{\today}

\begin{abstract}
\noindent Using Kolmogorov-Arnold Networks (KANs), we construct a holographic model informed by lattice QCD data. This neural network approach enables the derivation of an analytical solution for the deformation factor $w(r)$ and the determination of a constant $g$ related to the string tension. Within the KANs-based holographic framework, we further analyze heavy quark potentials under finite temperature and chemical potential conditions. Additionally, we calculate the drag force, jet quenching parameter, and diffusion coefficient of heavy quarks in this paper. Our findings demonstrate qualitative consistency with both experimental measurements and established phenomenological model.
\end{abstract}

\maketitle
{\it Introduction:} High-energy nuclear collisions conducted at facilities such as the Relativistic Heavy-Ion Collider (RHIC) and the Large Hadron Collider (LHC) provide a unique opportunity to study the properties of the quark-gluon plasma (QGP), a deconfined state of matter formed under extreme temperatures and densities. The QGP offers a window into the fundamental aspects of QCD, and understanding its dynamics is crucial for advancing our knowledge of strong interactions. However, the computational and theoretical resolution of QCD still faces many challenges in practical applications and complex environments. To address these challenges, the Anti-de Sitter/conformal field theory (AdS/CFT) correspondence provides a way to solve the problem. Through this dual relationship, strong interactions can be explored in higher-dimensional spaces \cite{Maldacena:1997re,Witten:1998zw,Gubser:1998bc}, offering new perspectives on the complex behavior of QCD. 

Experimental and theoretical studies have found that heavy quarks are an important tool in the study of finite-temperature QCD matter due to their unique behavior under extreme conditions \cite{Tang:2020hzi,ALICE:2024gqr,Andronic:2015wma,Luo:2017faz,Tang:2020ame,Liu:2013imm,Hattori:2016emy,Shou:2024uga,Ko:2016ioz,Averbeck:2013oga,Aarts:2007pk,Matsui:1986dk,QuarkoniumWorkingGroup:2004kpm,Rapp:2009my,Zheng:2024mep,Chen:2024iil,Chen:2024aom}. Under these conditions, the dissociation of heavy quark-antiquark pairs serves as a key indicator of strong interaction deconfinement, and thus it is critical to investigate the heavy quark potential within holographic QCD frameworks. Early work has tentatively revealed the characteristics of the holographic quark-antiquark pair potential energy \cite{Andreev:2006ct,Andreev:2006eh,Andreev:2006nw,He:2010bx,Colangelo:2010pe,DeWolfe:2010he,Li:2011hp,Fadafan:2011gm,Fadafan:2012qy,Cai:2012xh,Li:2012ay,Fang:2015ytf,Yang:2015aia,Zhang:2015faa,Ewerz:2016zsx,Chen:2017lsf,Arefeva:2018hyo,Bohra:2019ebj,Chen:2019rez,Zhou:2020ssi,Zhou:2021sdy,Chen:2020ath,Chen:2021gop,Guo:2024qiq,Bellantuono:2019xyy}. However, as the study progresses, understanding the dynamics of the potentials between quarks becomes essential to reveal the complexity of QCD matter.

Jets and heavy quark diffusion are among the most useful probes for investigating the properties of the QGP at different scales. Jets, produced from high-energy partons in collisions, undergo complex processes such as energy loss, medium-induced radiation, and medium response as they traverse the QGP. The substructure of jets, particularly observables like the Energy-Energy Correlator (EEC), provides detailed insights into the interaction between jets and the QGP. The EEC is highly sensitive to the angular distribution of energy within jets, revealing the interplay of mass effects, energy loss, and medium response. Studies have shown a clear flavor hierarchy in the EEC for both vacuum and QGP environments, driven by the mass effect of heavy quarks (e.g., charm and bottom quarks). By analyzing heavy-flavor jets, researchers can probe the mass dependence of jet substructure and jet-medium interactions, offering a deeper understanding of QGP dynamics \cite{Xing:2024yrb,Xing:2023ciw,JETSCAPE:2021ehl,Xie:2022ght,JET:2013cls,Majumder:2010qh,Wang:2024yag,Cai:2024mxf,Li:2025ugv}. In parallel, the diffusion of heavy quarks in the QGP is a critical aspect of understanding the hydrodynamic behavior of the plasma. Heavy quarks, such as charm and bottom quarks, are produced in the early stages of collisions and participate in the entire evolution of the QGP. The spatial diffusion coefficient, quantifies the momentum transfer from the QGP to heavy quarks and provides insights into the hydrodynamization process. Recent lattice QCD calculations with dynamical quarks have revealed that the heavy quark diffusion coefficient is significantly smaller than previous estimates from quenched lattice QCD and phenomenological models. This suggests that heavy quarks hydrodynamize very quickly in the QGP, highlighting the near-perfect fluidity of the medium \cite{Altenkort:2023oms,Casalderrey-Solana:2006fio,Caron-Huot:2007rwy,Altenkort:2020fgs,Brambilla:2020siz,Banerjee:2022gen,ALICE:2021rxa,Zhu:2019ujc}.

In recent years, machine learning techniques, especially multilayer perceptrons (MLPs), have shown unprecedented potential in solving complex scientific problems \cite{Ma:2023zfj,Zhou:2023tvv,Pang:2024kid,Ma:2023zfj,Li:2023eys,Pu:2023jae,Pang:2020ipp,Steinheimer:2019iso,Wang:2023yul,Huang:2025rfg,He:2023zin,He:2021uko,Wang:2020tgb,Guo:2023phd,10.1088/1674-1137/adb2fa,Kou:2024hzd}. According to the Universal Approximation Theorem, MLPs are able to approximate arbitrarily complex functions by increasing the number of neurons in the hidden layers, a feature that makes them perform well in solving partial differential equations (PDEs). Recently, Kolmogorov-Arnold Networks (KANs) have been proposed in Refs. \cite{Liu:2024swq,liu2024kan20kolmogorovarnoldnetworks}. Unlike traditional MLPs, KANs fundamentally eliminate the reliance on linear weight matrices by using learnable functions instead of fixed activation functions. For small-scale AI and scientific tasks, KANs may offer advantages in terms of accuracy and interpretability \cite{Erdmann:2024unt,Yu:2024isi}. When compared with conventional methods such as polynomial fitting, KANs may exhibit a difference. Polynomial fitting essentially performs data interpolation through linear combinations of basis functions, where the coefficients $\{a_n\}$ merely capture local data curvature without physical interpretability. 

With the rapid development of machine learning technology \cite{He:2023urp,Zhou:2018hsl,Pang:2019int,Wang:2020hji,Zhao:2021yjo,Du:2020pmp,Du:2021pqa,Jiang:2021gsw,Shi:2022vfr,Soma:2022qnv,Li:2022ozl,Yang:2022yfr,Soma:2023rls,Wang:2024dzc,Wang:2024bpl,Wang:2024ykk,Xu:2024tjp,Xiang:2024pkb,Zhou:2023pti,Zhou:2023dyd}, deep learning has received pioneering attention in the study of holographic QCD \cite{ALICE:2024rpm,Hashimoto:2018bnb,Hashimoto:2019bih,Akutagawa:2020yeo,Lu:2025tzs}. In addition, the combination of machine learning and holography has been deeply explored in a series of recent studies \cite{Hashimoto:2021ihd,Luo:2024iwf,Mansouri:2024uwc,Chen:2024ckb,Chen:2024mmd,Chang:2024ksq,Ahn:2024jkk,Cai:2024eqa,Hashimoto:2024yev,Ahn:2025tjp, Ahn:2024gjf, Lee:2025mti}. In contrast to traditional holographic models, this new approach first utilizes experimental or lattice QCD data to determine metrics and other model parameters with the help of machine learning. These acquired metrics are then applied to compute other physical QCD observables and are used as predictive outputs of the model. This interdisciplinary integration not only simplifies the complex computational process but also improves the prediction accuracy and reliability of the model, thereby opening up new directions in the study of strong interactions. \textcolor{black}{In our recent work \cite{Mansouri:2024uwc}, we utilize an emergent metric constructed by Neural Ordinary Differential Equations with QCD data of the chiral condensate to calculate real and imaginary potential of heavy quarkonium. Ref. \cite{Chen:2024epd} employs a machine learning-assisted Einstein-Maxwell-Dilaton (EMD) model, using automatic differentiation to determine six parameters based on the equation of state, to calculate the transport properties of heavy quarks. Ref. \cite{Luo:2024iwf} primarily focuses on comparing MLPs and KANs for the inverse problem of the heavy quark potential, confirming the validity of the constructed model at finite temperature and chemical potential. These works inspire us to incorporate more physical quantities into a single holographic model. In this study, we want to establish a connection between the heavy-quark potential and transport properties.}

This paper is organized as follows: Section \ref{sec1} will detail the calculation of the heavy quark potential using the Andreev-Zakharov model. In Section \ref{f}, we construct a holographic model based on KANs and calculate the heavy-quark potential at finite temperatures and chemical potentials. Our analysis will determine the critical temperature $T_c$ and provide an analytical solution for $w(r)$. In Section \ref{sec3}, the drag and jet quenching parameters, as well as the diffusion coefficients of heavy quarks at finite temperature and finite chemical potential, will be calculated for both regimes based on the analytic form of the function \( w(r) \). Finally, Section \ref{sec4} will summarize the main results and conclusions of the whole paper.

\section{Holographic heavy quark potential}
\label{sec1}
The Andreev-Zakharov model allows for an accurate description of the potential of heavy quarkonium  \cite{Andreev:2006ct,Andreev:2006nw} and exotic hadron states \cite{Andreev:2008tv,Andreev:2015riv,Andreev:2020xor,Andreev:2021bfg,Andreev:2021eyj,Andreev:2022qdu,Andreev:2024orz} by introducing an ad hoc deformation factor for the $\rm AdS_5$-RN black hole. The background metric can be expressed as
\begin{equation}ds^2=w(r)\frac{L_{AdS_5}}{r^2}\big[-f(r)dt^2+d\vec{x}^2+f^{-1}(r)dr^2\big].\end{equation}
\begin{equation}f(r)=1-\left(\frac{1}{r_h^4}+q^2r_h^2\right)r^4+q^2r^6.\end{equation}
\(w(r)\) is the deformation factor, which determines the deviation from conformality and the $AdS_5$ space radius $L_{AdS_5}$ is conventionally set to one, i.e., $L_{AdS_5}$=1. \(q\) is the black hole charge and \(r_h\) is the position of the black hole horizon. \textcolor{black}{$r$ is the fifth dimension of the spacetime.}  

In this paper,we can derive
\begin{equation}f(r)=1-\left(\frac{1}{r_h^4}+\frac{\mu^2}{r_h^2}\right)r^4+\frac{\mu^2}{r_h^4}r^6.
	\label{eq3}
\end{equation}
\begin{equation}T=\frac1{\pi r_h}\left(1-\frac12\mu^2r_h^2\right).
\label{eq4}
\end{equation}

The separation distance $L$ between the quark and antiquark is defined as
\begin{align}
L &= 2\int_0^{r_0}\partial_r x \, dr \nonumber\\
  &= 2\int_0^{r_0}\sqrt{\frac{\dfrac{f^2(r_0)w^2(r_0)}{r_0^4}}{\dfrac{f^2(r)f(r_0)w^2(r)}{r^4}-\dfrac{f^2(r_0)f(r)w^2(r_0)}{r_0^4}}}\,dr \label{eq5}
\end{align}
The heavy-quark potential can be written as
\begin{align}
E &= 2g \int_0^{r_0} \left( \frac{w(r)}{r^2} \sqrt{1 + f(r)(\partial_r x)^2} - \frac{w(0)}{r^2} - \frac{w'(0)}{r} \right) dr \nonumber\\
  &\quad - 2 \frac{g}{r_0} w(0) + 2g w'(0) \ln(r_0) \nonumber\\[2mm]
  &= 2g \int_0^{r_0} \Biggl( \frac{w(r)}{r^2} \sqrt{1 + \frac{\dfrac{f(r_0) w^2(r_0)}{r_0^4}}{\dfrac{f(r) w^2(r)}{r^4} - \dfrac{f(r_0) w^2(r_0)}{r_0^4}}} \nonumber\\
  &\qquad - \frac{w(0)}{r^2} - \frac{w'(0)}{r} \Biggr) dr - 2 \frac{g}{r_0} w(0) + 2g w'(0) \ln(r_0) \label{eq17}
\end{align}

$ r_0 $ (and $ r, r_h, r_s $) are dimensionless with $ L_{AdS_5} = 1 $; $\ln(r_0)$ and $ -2 r_0 g w(0) $ are UV divergence counterterms standard in holographic QCD used to cancel the power-law/logarithmic divergences of the integrand at $ r \to 0 $ and residual boundary divergences at $ r = r_0 $, ensuring a finite physical potential $ E $. The physical separation $ L $ is mapped from the dimensionless $ r_0 $ via Eq. (\ref{eq5}).

In the Andreev-Zakharov model, the parameter $g$ (related to the string tension) is fixed as \(g=0.176\), while the deformation factor \(w(r)\) takes the form \(w(r) = e^{sr^2}\) with \(s=0.45\). These values are determined from fits to the meson spectrum \cite{Andreev:2006vy, ParticleDataGroup:2004fcd} and the Cornell potential \cite{Andreev:2006ct, Eichten:1978tg}. At zero temperature, we only need to set \(f(r)=1\). In the next section, we will first use the KANs to construct the holographic model at vanishing temperature.

\section{Construction of a KANs-based holographic model}
\label{f}
\begin{figure}
	\centering
	\begin{minipage}[b]{0.47\textwidth}
		\includegraphics[width=1\textwidth]{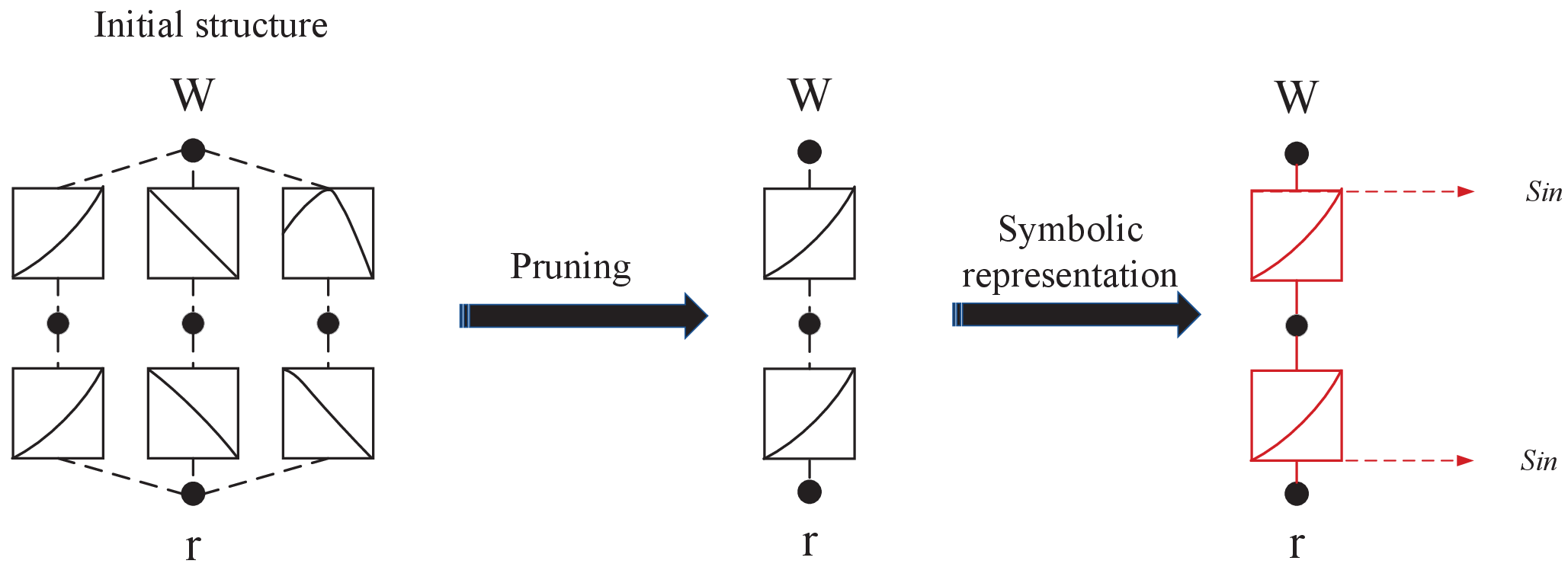}
	\end{minipage}
	
	\caption{A sketch of the KANs architecture used in our paper during the training progress.}
	\label{1}
\end{figure}

\begin{figure}
	\centering
	\begin{minipage}[b]{0.457\textwidth}
		\includegraphics[width=1\textwidth]{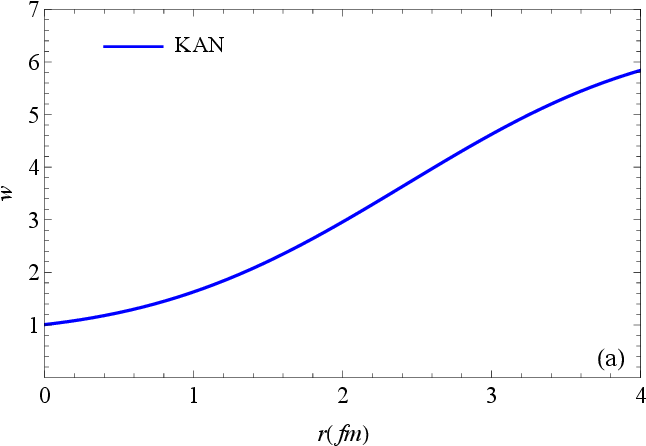}
	\end{minipage}
	\hfill
	\begin{minipage}[b]{0.47\textwidth}
		\includegraphics[width=1\textwidth]{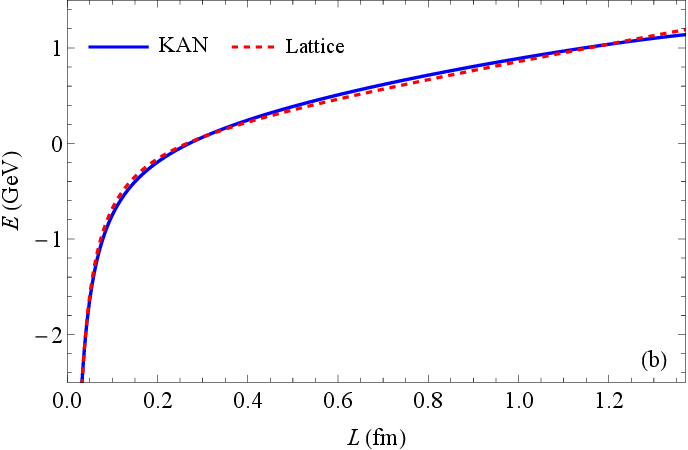}
	\end{minipage}
	
	\caption{$(a)$ The deformation factor $w(r)$ trained by KANs. $(b)$ Comparison of the heavy-quark potential from 2-flavor lattice QCD data with KANs predictions. The blue solid line represents the calculation results of  KANs, and the red dashed line represents the 2-flavor lattice data \cite{Kaczmarek:2005ui}.}
	\label{2}
\end{figure}

Based on KANs, we reconstruct the deformation factor \(w(r)\) in this section. We assume that \(w(r)\) is a specific function derived from the lattice data, with the string tension \(g\) used as a free parameter. The loss function integrates four essential components to achieve both physical fidelity and mathematical robustness. First, it incorporates a deviation, specifically the mean absolute error, which quantifies the discrepancy between model predictions of the heavy-quark potential calculated by Eq. (\ref{eq17}) and the target values from lattice QCD data of the static heavy-quark potential \cite{Kaczmarek:2005ui}, using the Cornell potential \( E = 0.8404L - \frac{0.0866}{L} + 0.1033 \) as the target value to achieve a perfect fit to the lattice data. Second, regularization constraints are imposed through specialized mathematical mechanisms to prevent overfitting and preserve generalization capabilities. Third, by calculating the absolute difference between \( w(0) \) and 1 and minimizing it, we can ensure that the ultraviolet boundary condition rigorously enforces \( w(0) \rightarrow 1 \). This is a fundamental requirement that ensures the metric asymptotically transitions to \( AdS_5 \) spacetime at short distances. Finally, by comparing the radial point \( \frac{w(r_n)}{r_n^2} \) with the boundary value \( \frac{w(r_0)}{r_0^2} \), a custom penalty is applied to impose a radial physical constraint while strictly maintaining the increasing behavior of \( w(r) \) across its domain, thereby eliminating unphysical imaginary solutions in subsequent computations. The structure of the KANs we designed is shown in Fig. \ref{1}, we performed a reconstruction with the parameter \(g=0.2573\) related to the string tension. The trained function \(w(r)\) is $w(r) = 2.94 \sin\left(1.24 \sin\left(0.47 r + 1.94\right) - 9.55\right) + 3.53$, which is substituted into Eq. (\ref{eq17}) for potential energy calculations. These calculations are combined with the lattice QCD data \cite{Kaczmarek:2005ui}, and the results are shown in Fig. \ref{2}.

From Fig. \ref{2} ($a$), it can be clearly observed that the reconstructed function satisfies the boundary condition \( w(0) \) $\rightarrow$ 1 and the function $w(r)$ is increasing with $r$. Fig. \ref{2} ($b$) illustrates the fitting performance of the neural network to the training dataset. These curves clearly demonstrate the excellent ability of the network in modeling the function \( E(L) \) with an accuracy that highly matches the theoretical expectation.  This result further supports the validity of the model, verifies the reliability of the KANs.

\begin{figure}
	\centering
	\begin{minipage}[b]{0.47\textwidth}
		\includegraphics[width=1\textwidth]{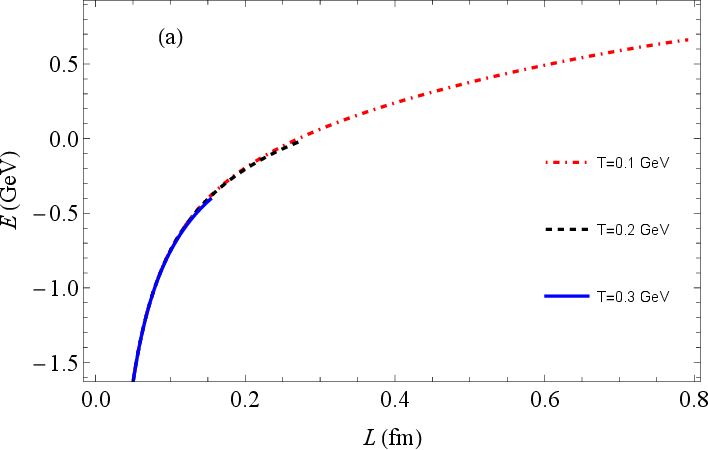}
	\end{minipage}
	\hfill
	\begin{minipage}[b]{0.47\textwidth}
		\includegraphics[width=1\textwidth]{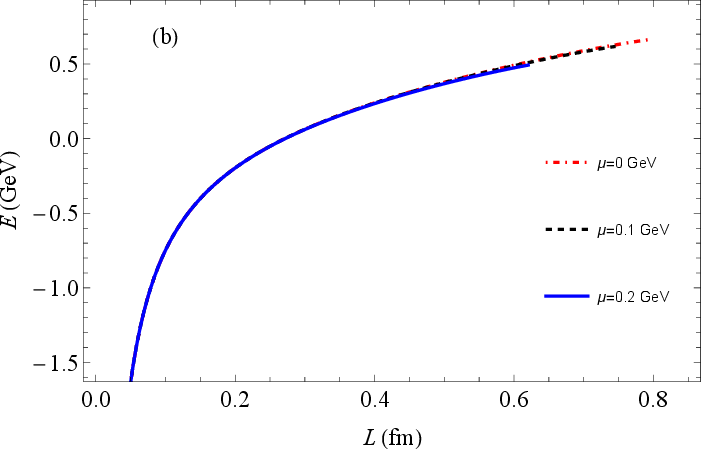}
	\end{minipage}
	
	\caption{(\(a\)) Potential energy \(E\) of quark-antiquark pairs as a function of quark separation distance \(L\) at different temperatures $T$ when chemical potential $\mu$=0. (\(b\)) Variation of quark separation distance \(L\) with potential energy \(E\) for different chemical potentials at fixed temperature \(T=0.1\). The unit of \(E\) is GeV, the unit of \(L\) is fm, the unit of $\mu$ is GeV, the unit of \(T\) is GeV.}
	\label{h}
\end{figure}

In the subsequent stage of this study, we obtain an analytical solution of \(w(r)\). From this solution, we calculate the heavy quark potential under finite temperature and chemical potential conditions by adding the function \(f(r)\). Fig. \ref{h} (\(a\)) shows that the linear component of the potential energy exhibits a decaying trend under finite temperature conditions. This decay may be attributed to the temperature-induced shielding effect. In contrast, the Coulombic component of the potential energy shows remarkable robustness to temperature changes and is almost unaffected by temperature increase. As the temperature increases, the coupling strength of the strong interactions weakens, leading to weaker inter-quark binding and the potential energy tends to vanish on a smaller spatial scale. Ref. \cite{Bala:2021fkm} employs lattice QCD simulations to analyze static quark-antiquark interactions at finite temperature using four complementary methods: spectral model fits, HTL-inspired fits, Padé rational approximation, and Bayesian reconstruction (BR). Among these, only the HTL-inspired fits exhibit significant temperature dependence.  Our results are consistent with the other three methods, indicating that the screening effect is not significant at finite temperature.  Fig. \ref{h} (\(b\)) shows a similar trend. However, by comparison (\(a\)), we find that the chemical potential affects the potential energy to a significantly lesser extent than temperature does. These observations reveal the differential effects of temperature and chemical potential on the strong interaction potential energy. 

\begin{figure}
	\centering
	\begin{minipage}[b]{0.47\textwidth}
		\includegraphics[width=1\textwidth]{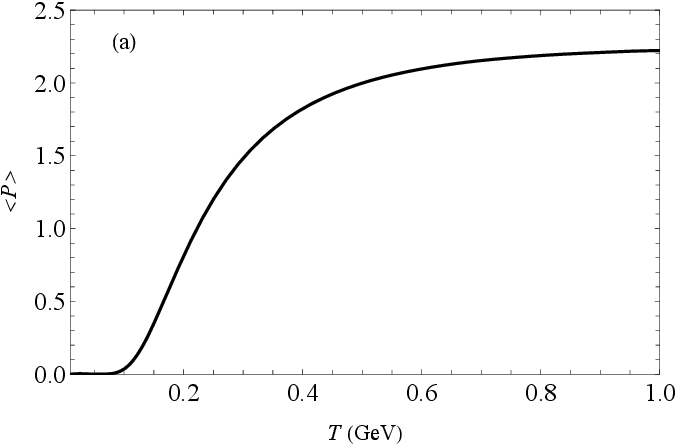}
	\end{minipage}
	\hfill
	\begin{minipage}[b]{0.47\textwidth}
		\includegraphics[width=1\textwidth]{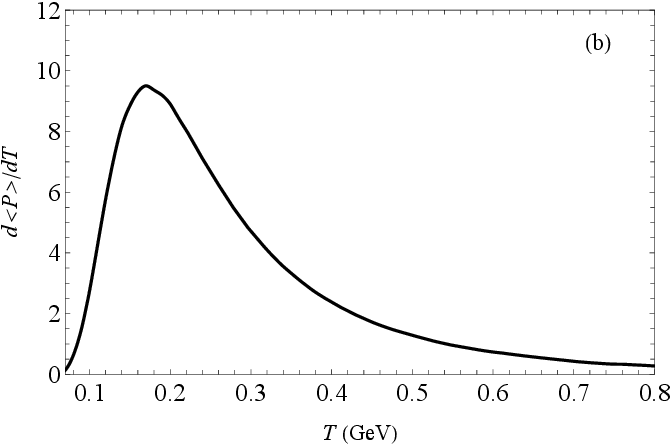}
	\end{minipage}
	
	\caption{$(a)$ The relationship between the Polyakov loop expectation value $<P>$ and temperature $T$. $(b)$ The relationship between the slope of the Polyakov loop expectation value $\frac{d<P>}{dT}$ and temperature $T$. The unit of \(T\) is GeV.}
	\label{n}
\end{figure}

At larger distance scales, the interaction between quarks and antiquarks is significantly weakened so that they behave as free particles. This phenomenon suggests that when the separation distance between quarks and antiquarks is large enough, they can be regarded as independently existing states and are no longer bound by strong interactions. The expectation value of the Polyakov loop can be defined as \cite{Andreev:2006nw,Colangelo:2010pe}
\begin{equation}
	<P> = \exp\left\{-\frac{1}{2T} E(r = \infty, T)\right\}.
	\label{eq19}
\end{equation}
With this choice, the Polyakov loop expectation value takes the form shown in Fig \ref{n} $(a)$. From Fig. \ref{n} $(b)$, it can be clearly seen that at $T_c = 0.17$ GeV, the slope of the Polyakov loop expectation value is the largest.

\begin{figure}
	\centering
	\begin{minipage}[b]{0.47\textwidth}
		\includegraphics[width=1\textwidth]{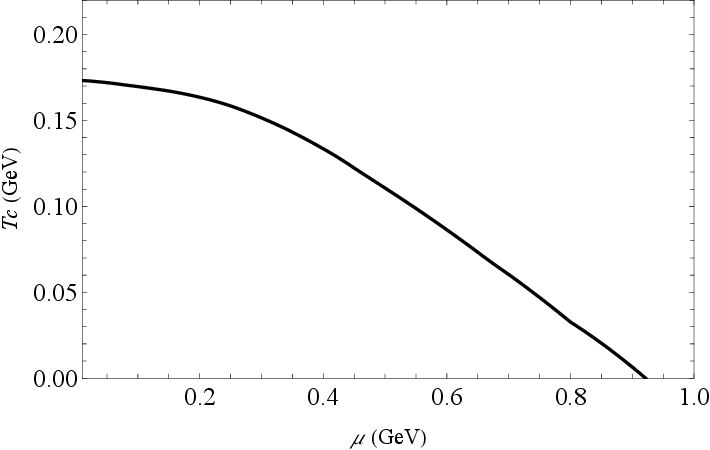}
	\end{minipage}
	
	\caption{Deconfinement transition line in the $\mu$ - $T_c$ plane. \(T\) and $\mu$ are in units of GeV.}
	\label{T}
\end{figure}

As shown in Fig. \ref{T}, $w(r)$ is monotonically increasing and satisfies $w(0) \rightarrow 1$, ensuring the UV asymptotic $AdS_5$ geometry. At finite $\mu$ and $T$, the blackening factor $f(r)$ in Eq. (\ref{eq3})  together with $w(r)$  determines the horizon position $r_h$ and the charge $q$, which are related to $\mu$ via Eq. (\ref{eq4}). The resulting phase boundary exhibits a monotonic decrease of $T_c$ with increasing $\mu$, and vanishes at $\mu \approx 0.93$~GeV. The picture qualitatively agrees with the diagram obtained by, e.g., Nambu-Jona-Lasinio \cite{BUBALLA_2005} and other effective models \cite{Schaefer_2005}.

\section{Transport properties of QGP}
\label{sec3}
\begin{figure}
	\centering
	\begin{minipage}[b]{0.47\textwidth}
		\includegraphics[width=1\textwidth]{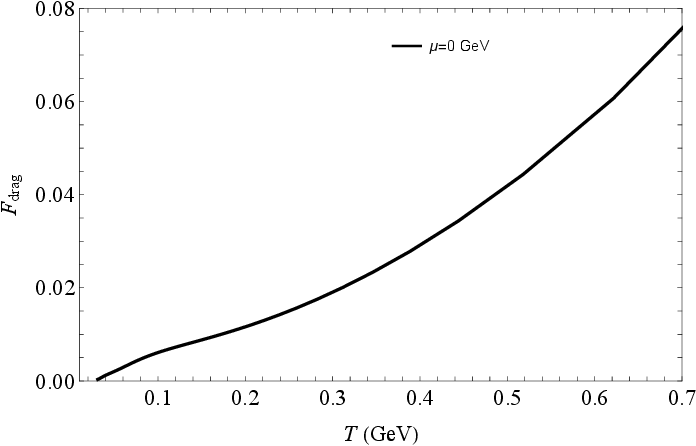}
	\end{minipage}
	
	\caption{Variation of drag force with temperature when quarks are in \(v=0.3\) and $\mu=0$. \(T\) and $\mu$ are in units of GeV.}
	\label{5}
\end{figure}

In this section, we extend the calculation of analytic solutions of the function \( w(r) \) with string tension \( g \) to the drag force, diffusion coefficient, and jet quenching parameter of heavy quarks at finite temperature and finite chemical potential. The holographic formalism for relating the drag force of heavy quarks to the spatial string tension in deformed AdS spacetimes has been well established in ref \cite{Andreev:2017bvr}, which provides the foundational framework for our transport property calculations with the KANs-reconstructed $w(r)$. To facilitate the study of holographic probes, we define \( A_s(r) = \frac{1}{2} \log(w(r)) \). Following Ref. \cite{Chen:2024epd,Gubser:2006bz,Herzog:2006gh,Liu:2006ug}, the drag force can be obtained by:
\begin{equation}
	F_{\text{drag}} = \frac{dp}{dt} = \frac{dE}{dx} = -\frac{1}{2\pi \alpha'} \frac{e^{2A_s(r_s)} v}{r_s^2},
	\label{eq21}
\end{equation}
where $r_s$ satisfies $f(r_s)-v^2=0$. In this context, we first need to solve Eq. (\ref{eq5}) numerically to obtain \(r\), and subsequently employ Eq. (\ref{eq19}) to compute the drag force. In Fig. \ref{5}, we show the variation of drag force with temperature at vanishing chemical potential. As can be seen from the figure, the drag force increases significantly with increasing temperature.

\begin{figure}
	\centering
	\begin{minipage}[b]{0.47\textwidth}
		\includegraphics[width=1\textwidth]{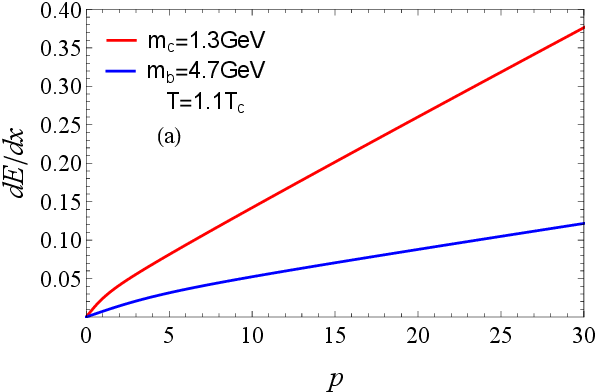}
	\end{minipage}
	\hfill
	\begin{minipage}[b]{0.463\textwidth}
		\includegraphics[width=1\textwidth]{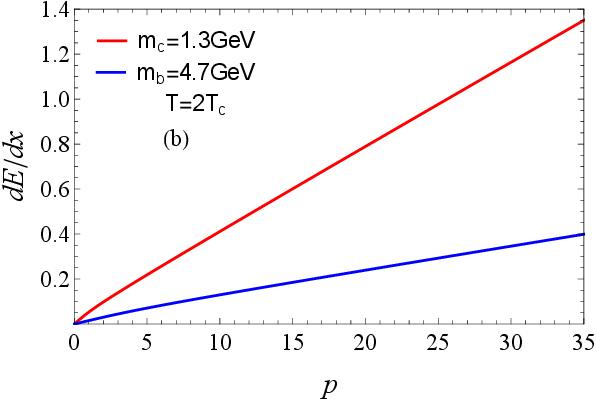}
	\end{minipage}
	\\
	\caption{Variation of energy loss with momentum $p$ for bottom (\(m_b=4.7\) GeV) and charm quark (\(m_c=1.3\) GeV). (\(a\)) Variation of energy loss with momentum \(p\) at temperature \(T=1.1T_c\). (\(b\)) Energy loss versus momentum at temperature \(T=2T_c\).}
	\label{6}
\end{figure}

According to Eq. (\ref{eq21}) can be deduced that the energy loss is equal to the drag force, which allows us to plot the relationship between energy loss and momentum in different systems. We have

\begin{equation}
	\frac{dE}{dx}= -\frac{e^{2A_s(r_s)}\sqrt{1 - v^2}}{\pi^2 T^2 r_s^2} \frac{\pi T^2 \sqrt{\lambda}}{2m} p.
\end{equation}
Fig. \ref{6} illustrates the energy loss of bottom quark (\(m_b=4.7\) GeV) and charm quark (\(m_c=1.3\) GeV) at vanishing chemical potential. From Fig. \ref{6}, it is clear that the energy loss increases with momentum. In addition, higher temperatures lead to an increase in energy loss. The qualitative results are similar to Refs. \cite{Guo:2024mgh,Du:2024riq,Chen:2024epd}.

\begin{figure}
	\centering
	\begin{minipage}[b]{0.47\textwidth}
		\includegraphics[width=1\textwidth]{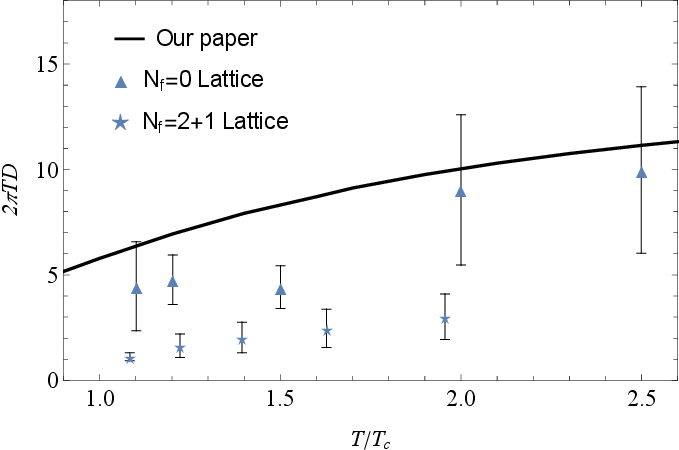}
	\end{minipage}
	
	\caption{The figure shows the normalized diffusion coefficient versus $T/T_c$. The unit of \(T\) is GeV.}
	\label{7}
\end{figure}

Next we proceed to study the diffusion coefficient, which in the AdS/Schwarzchild context can be written as \cite{Chen:2024epd,Gubser:2006bz,Herzog:2006gh,Liu:2006ug}
\begin{equation}
	D = \frac{T}{m} t = \frac{2}{\pi T g} \frac{\pi^2 T^2r_s^2}{e^{2A_s(r_s)} \sqrt{1 - v^2}}.
	\label{eq22}
\end{equation}
According to Eq. (\ref{eq22}), we calculate the heavy-quark diffusion coefficient normalized by 2$\pi T$, as shown in Fig. \ref{7}. As can be seen from the figure, the diffusion coefficient of heavy quarks gradually increases with increasing temperature, and this behavior is consistent with reference \cite{Altenkort:2023oms}, indicating that the analytic solution of \(w(r)\) obtained by KANs is reliable. Besides, the qualitative behavior of diffusion coefficient is consistent with Ref. \cite{Andreev:2017bvr}.

\begin{figure}
	\centering
	\begin{minipage}[b]{0.47\textwidth}
		\includegraphics[width=1\textwidth]{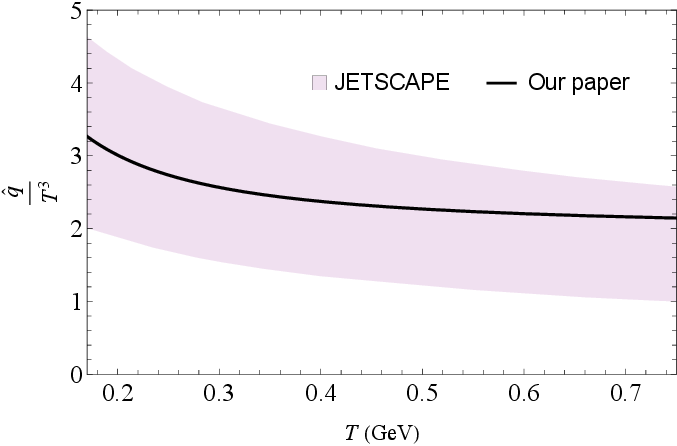}
	\end{minipage}
	
	\caption{Relationship between $\hat{q} / T^3$ and temperature \(T\) at zero chemical potential. The unit of \(T\) is GeV.}
	\label{8}
\end{figure}

\begin{figure}
	\centering
	\begin{minipage}[b]{0.47\textwidth}
		\includegraphics[width=1\textwidth]{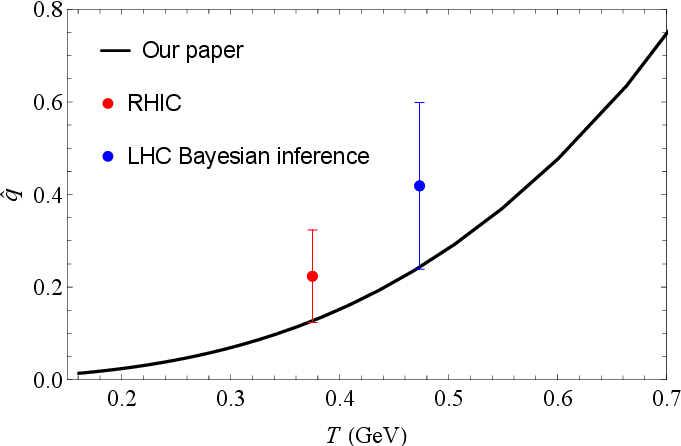}
	\end{minipage}
	
	\caption{The relationship between the jet quenching parameter and temperature in the holographic model at zero chemical potential. Error bars represent experimental values from RHIC and LHC \cite{JET:2013cls}. The unit of \(T\) is GeV.}
	\label{9}
\end{figure}

Now we turn to the study of jet quenching parameter and we obtain the following expression for the jet quenching parameter in the holographic model \cite{Chen:2024epd}
\begin{equation}
	\hat{q} = \frac{g}{\pi a_0}.
	\label{eq23}
\end{equation}
Here \(a_0\) is defined
\begin{equation}
	a_0 = \int_{0}^{r_h} \frac{dr \, r^2 L^{-2} e^{-2 A_s(r)}}{\sqrt{f(r)(1 - f(r))}}.
	\label{eq24}
\end{equation}
To obtain the jet quenching parameters in the holographic QCD model, we numerically solved using Eq. (\ref{eq23}), comparing $\hat{q} / T^3$ as a function of temperature, as shown in Fig. \ref{8}. From Fig. \ref{9}, it can be observed that temperature leads to an enhancement of the jet quenching parameter, indicating that in the considered model, a denser or hotter medium results in increased energy loss, which aligns with the physical intuition that jets passing through a higher-temperature (i.e., higher-density or more particle-rich) medium encounter more scattering centers and therefore experience greater energy loss. This behavior is in qualitative agreement with \cite{Zhu:2023aaq}, where the jet quenching parameter in Einstein-Maxwell-dilaton holographic model (with finite $\mu$ and magnetic field) exhibits similar temperature-induced enhancement, and the peak value shifts with background field strength supporting the reliability of our KANs-based model in capturing medium effects on jet energy loss. The results of our model calculations are consistent with the experimental results of RHIC and LHC \cite{JET:2013cls}.


\section{ Summary} 	
\label{sec4}
In this study, we employ the KANs to extract information from QCD data to construct a holographic model. In order to verify the validity of the reconstruction results, we first apply the obtained function \(w(r)\) to the heavy-quark potential and compare with lattice data. The results show that KANs exhibit effectiveness in solving inverse problems. It is worth emphasizing that KANs have the ability to provide analytical solutions. Moreover, we further examine the heavy quark potential and its relationship at finite temperature and chemical potential. 

In addition, based on the constructed function \(w(r)\), we study the relationship between the drag force of heavy quarks, the diffusion coefficient, and the jet quenching parameter under the conditions of finite temperature and finite chemical potential, which reveals the accuracy of the KANs-based holographic model. Our results on the temperature/chemical potential dependence of transport properties are consistent with recent findings in \cite{Arefeva:2025uym}, which emphasize the role of $\hat{q}$ as a phase transition probe in anisotropic and magnetized QGP highlighting the potential of our KANs-based framework for exploring complex QCD matter scenarios. Compared to MLPs, we identify two key advantages of KANs in our paper. First, the fitting time of KANs is significantly shorter than that of MLPs, as demonstrated by our implementation available on GitHub for direct runtime comparison. For example, KANs take only 851 seconds to reach a loss function value of 0.0328, whereas MLPs require 15,991 seconds to reach a loss of 0.0847. This clearly illustrates the superior training efficiency of KANs compared to MLPs. Second, our results confirm KAN's ability to reveal the compositional structure of functions, highlighting its interpretability advantage over traditional approaches.

The findings of this study provide a preliminary reference for using machine learning methods to address physics problems, as well as suggest some initial exploratory directions for future research. These new directions include an in-depth exploration of the interactions between different physical fields, and the use of holographic models in combination with machine learning methods. 

Finally, we want to emphasize that our results focus on qualitative behavior. On the one hand, we can develop a holographic model derived from the Einstein equation using KANs. On the other hand, this work could inspire future efforts to incorporate additional data (e.g., meson spectra, equations of state) into the holographic model. Ultimately, this may enable us to build a comprehensive framework for describing broader aspects of QCD physics.

\section*{Acknowledgments}
 This work is supported by National Natural Science Foundation of China (NSFC) Grants No. 12405154, 12175100, and the European Union -- Next Generation EU through the research grant number P2022Z4P4B ``SOPHYA - Sustainable Optimised PHYsics Algorithms: fundamental physics to build an advanced society'' under the program PRIN 2022 PNRR of the Italian Ministero dell'Universit\`a e Ricerca (MUR).
 
\section*{References}

\bibliography{ref}
\end{document}